# On a periodicity measure and superoscillations

Nehemia Schwartz[1] and Moshe Schwartz[2]

[1] Department of Physics, Bar- Ilan University, 5290 Ramat-Gan , Israel

[2] School of Physics and Astronomy, Raymond and Beverly Sackler Faculty of Exact Sciences, Tel Aviv University, Tel Aviv 69978, Israel

**Abstract** -The phenomenon of superoscillation, where band limited signals can oscillate over some time period with a frequency higher than the band limit, is not only very interesting but it also seems to offer many practical applications. The first reason is that the superoscillation frequency can be exploited to perform tasks beyond the limits imposed by the lower bandwidth of the signal. The second reason is that it is generic and applies to any wave form, be it optical, electrical, sonic, or quantum mechanical. For practical applications, it is important to overcome two problems. The first problem is that an overwhelming proportion of the energy goes into the non superoscillating part of the signal. The second problem is the control of the shape of the superoscillating part of the signal. The first problem has been recently addressed by optimization of the super oscillation yield, the ratio of the energy in the superoscillations to the total energy of the signal. The second problem may arise when the superoscillation, is to mimic a high frequency purely perodic signal. This may be required, for example, when a superoscillating force is to drive a harmonic oscillator at a high resonance frequency. In this paper the degree of periodicity of a signal is defined and applied to some yield optimized superoscillating signals.

Super oscillatory functions are band limited functions that oscillate with a frequency larger than its maximal Fourier component over a limited time space domain.
A number of examples have been given in the past for such functions with very interesting applications to Quantum Mechanics [1-7] , Optics and super resolution [8-11] superdirectivity and supergain [12,13] and signal processing [14-18]. Superoscillations offer at first sight intriguing technological and scientific applications, due to the mere fact that low frequency oscillations combine over limited intervals to produce high frequency signals. It is well known, however that superoscillations exist in limited intervals of time (or regions of space, depending on the actual problem) and that the amplitude of the super oscillation in those regions is extremely small compared to typical values of the function in non-oscillating regions [2]. It is generically so small, that any hope of practical application of super oscillating functions depends on tailoring the functions carefully to reduce that effect as much as possible. The tailoring of the function has to include also the frequency of the superoscillations. The latter goal was achieved by forcing the function to interpolate through a set of prescribed points and the first goal was achieved by optimization of the signal with respect to the choice of the Fourier coefficients of the low frequency components of which the signal consists. The first optimization of that nature was done by Ferreira and Kempff [18] who minimize the energy of the signal, $E = \int_{-\infty}^{\infty} f^2(t)dt$, with respect to the choice of the band limited Fourier transform of the signal subject to the interpolation constraints. A more recent natural step forward [22] was to optimize the superoscillation yield, which is the ratio between the energy in the interval, $(-\alpha, \alpha)$ where the signal is constrained to oscillate with frequency, $\omega$, which is larger than the band cut-off, and the total energy,

$$Y(\omega, a) = \frac{\int_{-\alpha}^{\alpha} f^2(t)dt}{\int_{-\infty}^{\infty} f^2(t)dt}, \tag{1}$$

To be more specific about the shape problem consider a periodic signal of the form

$$f(t) = \frac{A_0}{\sqrt{2\pi}} + \sum_{m=1}^{N} \frac{A_m}{\sqrt{\pi}} \cos(mt) \tag{2}$$

Such that by proper choice of the Fourier coefficients, one would like $f(t)$ to mimic the function $\cos(\omega t)$ in the interval $(-\alpha, \alpha)$. This is achieved by forcing the interpolation $f(t_j) = (-1)^j$ at the points $t_j = \alpha j / (M-1)$, where $j = 0, \ldots, M-1$. This choice enforces an oscillation of frequency, $\omega = \frac{\pi(M-1)}{\alpha}$. For $M < N+1$ not only the constraints can be met but also the yield can be optimized in the subspace of unconstrained combinations of the Fourier coefficients $\{A_m\}$. Figure 1 demonstrates the superoscillatory part of a yield optimized low frequency signal with $N = 10$, $\alpha = 1$ and $M = 6$. It is evident that although the function interpolates through the required points and superoscillates, it does not really resemble the function it was intended to mimic.

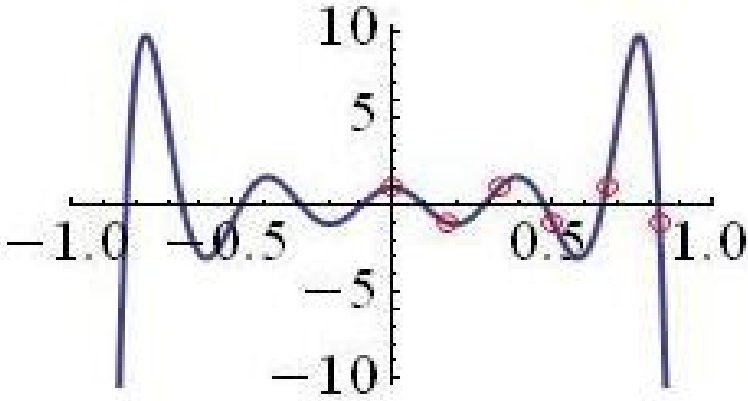

Figure 1: The superoscillatory interval of a yield optimized low frequency signal with $N=10$, $\alpha=1$ and $M=6$. The circles indicate the prescribed points through which the function is forced to interpolate.

In fact, it is not even periodic. This situation is generic. We cannot really expect any signal, which is a combination of low frequency oscillations, to be strictly periodic in its superoscillatory interval. Since we may be interested to have it mimic a strictly periodic function, it is important to define a figure of merit that will serve as a measure for the departure of a function in a given interval from a function which is strictly periodic which has a number of periods in that interval, which the superoscillatory signal is supposed to mimic

Let $g(t)$ be a continuous function defined, without loss of generality, in the interval $(-a,a)$, the purpose is to obtain a measure of the departure of $g$ from a periodic function with period $2\pi/\omega < 2a$. Now, $g(t)$ can be expressed in the interval $(\tau, \tau + 2\pi/\omega)$ as

$$g(t) = \sum_{n=-\infty}^{\infty} A_n(\tau,\omega)\exp(in\omega t), \qquad (3)$$

where

$$A_n(\tau,\omega) = \frac{\omega}{2\pi} \int_{\tau}^{\tau+2\pi/\omega} g(t)\exp(-in\omega t)dt. \qquad (4)$$

(Only at the points $\tau$ and $\tau + 2\pi/\omega$, the sum on the right hand side of equation (3) may not converge to its left hand side.) If the function $g(t)$ was periodic in the interval $(-a,a)$ with period $2\pi/\omega$, then as long as $\tau \geq -a$ and $\tau + 2\pi/\omega \leq a$, the Fourier coefficients would not depend on $\tau$. A figure of merit that measures the departure of $g(t)$ can be obtained in the following way. First averages of the Fourier coefficients over the interval are defined,

$$\overline{A}_n^{(a)}(\omega) = \frac{1}{(2a-\Delta)} \int_{-a}^{a-\Delta} d\tau A_n(\tau,\omega), \qquad (5)$$

where $\Delta = 2a - \frac{\pi}{\omega}\text{int}(\frac{a\omega}{\pi})$. Correspondingly the fluctuations in the Fourier coefficients are

$$\left\langle |\delta A_n^{(a)}(\omega)|^2 \right\rangle = \frac{1}{(2a-\Delta)} \{ \int_{-a}^{a-\Delta} d\tau |\delta A_n(\tau,\omega)|^2, \qquad (6)$$

where $\delta A_n(\tau,\omega) = A_n(\tau,\omega) - \overline{A}_n^{(a)}(\omega)$. These can be combined to define the degree of $2\pi/\omega$ periodicity of the function $g$ in the interval $(-a,a)$.

$$\eta\{g;a,\omega\} = 1 - \frac{\sum_{n=-\infty}^{\infty} \left\langle |\delta A_n^{(\alpha)}(\omega)|^2 \right\rangle}{\sum_{n=-\infty}^{\infty} \left\langle |A_n^{(\alpha)}(\omega)|^2 \right\rangle}. \qquad (7)$$

By construction, $0 \leq \eta \leq 1$ and the value $1$ is attained for fully periodic functions with period $T = 2\pi/\omega$. The expected period is to be determined just by observation. This can be done, for example, by counting the number of major peaks and troughs in the interval. Therefore it is very important to choose $a\omega/\pi$ to be considerably larger than 1. Otherwise there are no oscillation in the interval and $\eta$ which depends very strongly on that ratio, will be close to 1, just because all the $\tau$ dependent Fourier coefficients will involve integrals on almost overlapping intervals. In any case it should be remembered that the degree of periodicity depends strongly on $a\omega/\pi$ even when that parameter is larger than one.

To get a feeling for the periodicity measure, we consider the function

$$g(t;c) = \cos[t(1+c\cos t)]. \qquad (8)$$

In figure 1 we present $g$ as a function of $t$. It is clear that the periodicity ($2\pi$) of the function which is exact at $c=0$ is deteriorates as $c$ is increased.

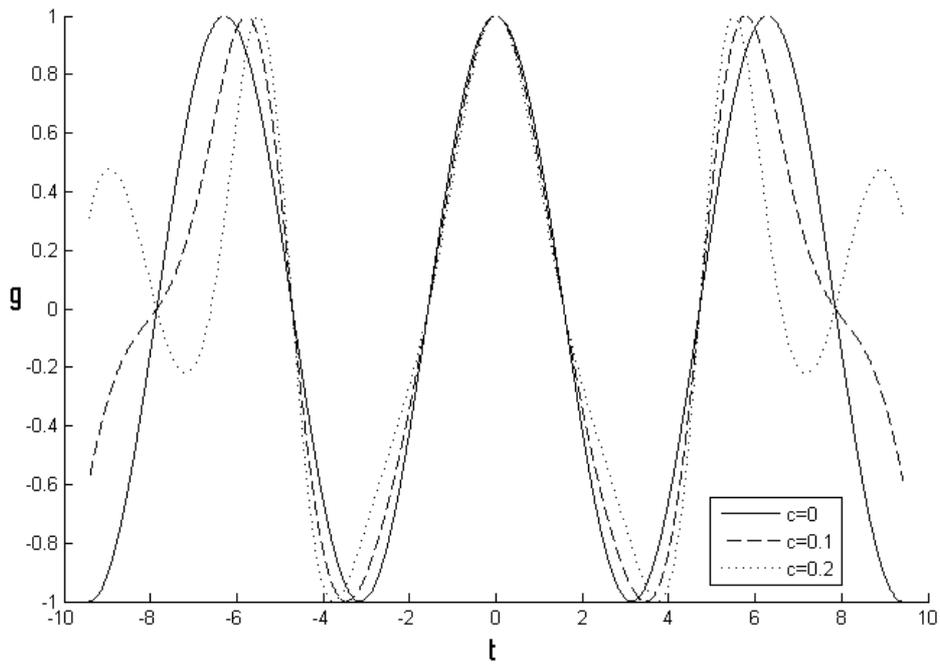

Fig. 2 The function $g(t;c)$ presented as a function of $t$ for three values of $c$.

Next we present two graphs of $\eta$ as a function of $c$. The first gives $\eta[2\pi,1]$. (We have omitted the dependence on $g$, which is clear.) The second is of $\eta[3\pi,1]$.

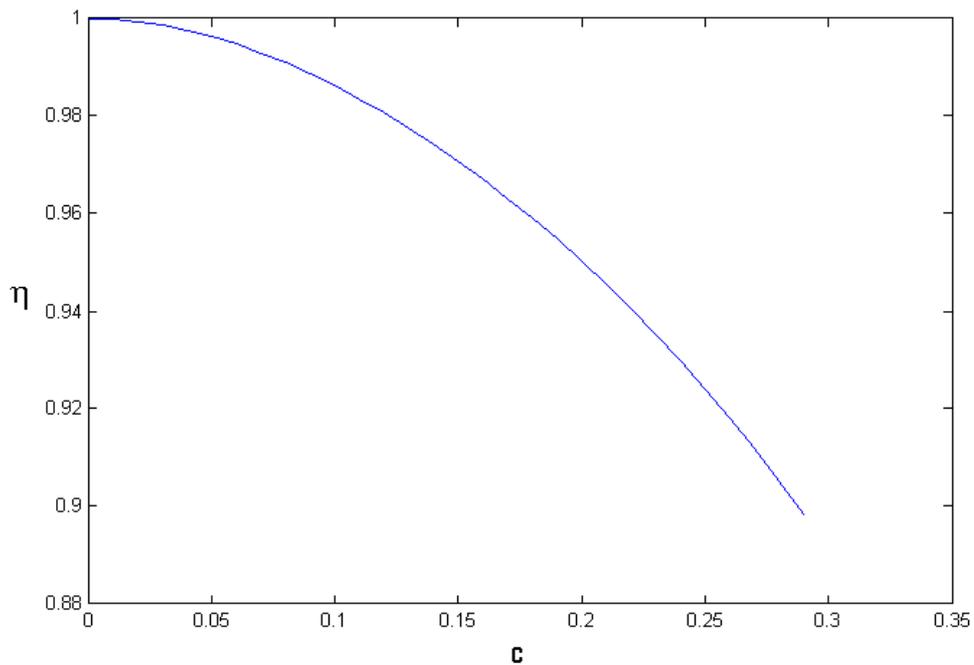

Fig. 3 $\eta$ vs. $c$ for $\omega=1$ and $a=2\pi$.

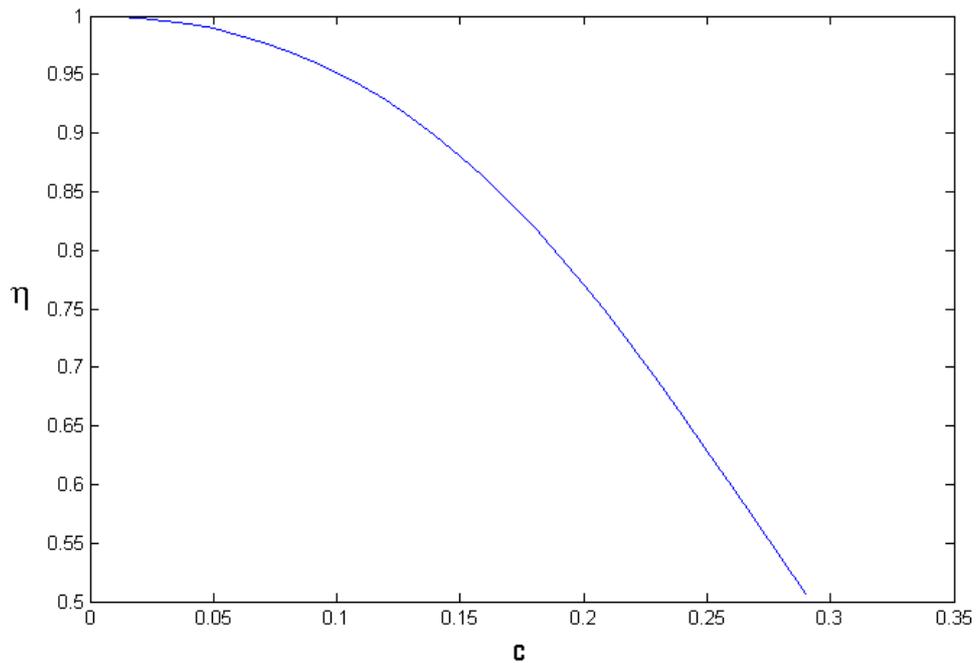

Fig. 4 $\eta$ vs. $c$ for $\omega=1$ and $a=3\pi$.

Note that although the shape of the last two graphs is similar, the values attained by $\eta$ for the larger interval are considerably smaller than those attained when the function is sampled over a smaller interval.

We have considered next the degree of periodicity for five superoscillatory signals characterized by $\alpha=1$, $M/N=2/3$ and bandwidth which is determined by $N=3K$ with $K=2,...,6$. Just to give an impression of the signals we consider, we present in the following two of those signals, the signals with the lowest and highest frequency among the five.

We start with the function characterized by $N=6$, $M=4$. For the benefit of the reader who would like to repeat the calculations we give the Fourier coefficients of the low frequency oscillations.

A[0]=-41.47083124305578;
A[1]=-20.32622763681456;
A[2]=46.37700031886999;
A[3]=52.21194270427686;
A[4]=-21.97868165245421;
A[5]=-73.5921499178442;
A[6]=48.40487602827931;

The corresponding function is presented below.

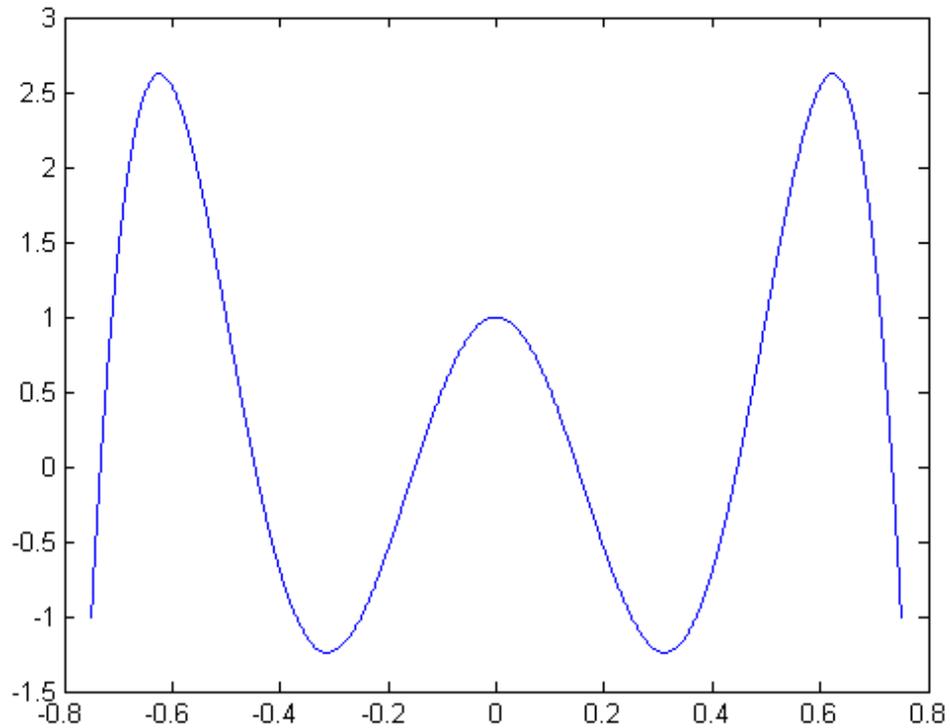

Fig. 5 The yield optimized superoscillation with $N=6, M=4$.

The second example is of optimized superoscillation with $N=18, M=12$.

```
A[0]=-1.903888878073059e13;
A[1]=1.180132233073902e13;
A[2]=1.663751956480845e13;
A[3]=-2.699745555272258e13;
A[4]=8.386654885278459e12;
A[5]=1.939746829024658e13;
A[6]=-2.896973240841378e13;
A[7]=1.255039615978049e13;
A[8]=1.488073774636226e13;
A[9]=-3.362069665948907e13;
A[10]=3.576662494942123e13;
A[11]=-2.643277912244821e13;
A[12]=1.476651494692829e13;
A[13]=-6.383635359962782e12;
A[14]=2.128038908223024e12;
A[15]=-5.335251951977357e11;
A[16]=9.536792743266927e10;
A[17]=-1.089456678031599e10;
A[18]=6.005189068534335e8;
```

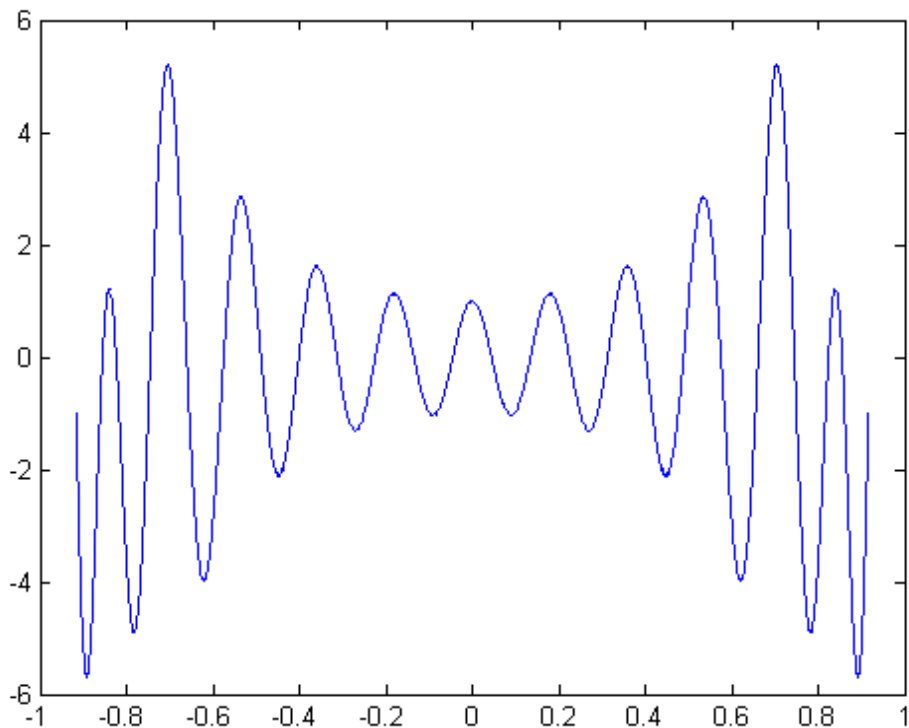

Fig. 6 The yield optimized superoscillation with $N=18, M=12$.

What do we learn from the above?

From the coefficients of the Fourier components we see that the phenomenon of superoscillation is short of miraculous. The absolute size of the coefficients increases very steeply with $N$ ,yet their signs that look almost random conspire to generate a region where cancellations occur and the function is of order one.

We see that when see that the higher the frequency, the more are the most internal oscillations "protected" and look like real oscillations. This is true also for the non-optimal but analytic super oscillations introduced many years ago by Y. Aharonov [2] ,

$$f_n(t) = (\cos(t/n) + i\omega \sin(t/n))^n ,\qquad(14)$$

That behaves as $e^{i\omega t}$ for $t$ in an interval of order $\sqrt{n}$ around the origin.

Let us see if our visual impression is supported by the degree of periodicity. In the following we present two graphs that cover the degree periodicity of all five aignals as a function of

$$\frac{2a}{T} = \frac{2\pi a}{\alpha(M-1)}. \tag{15}$$

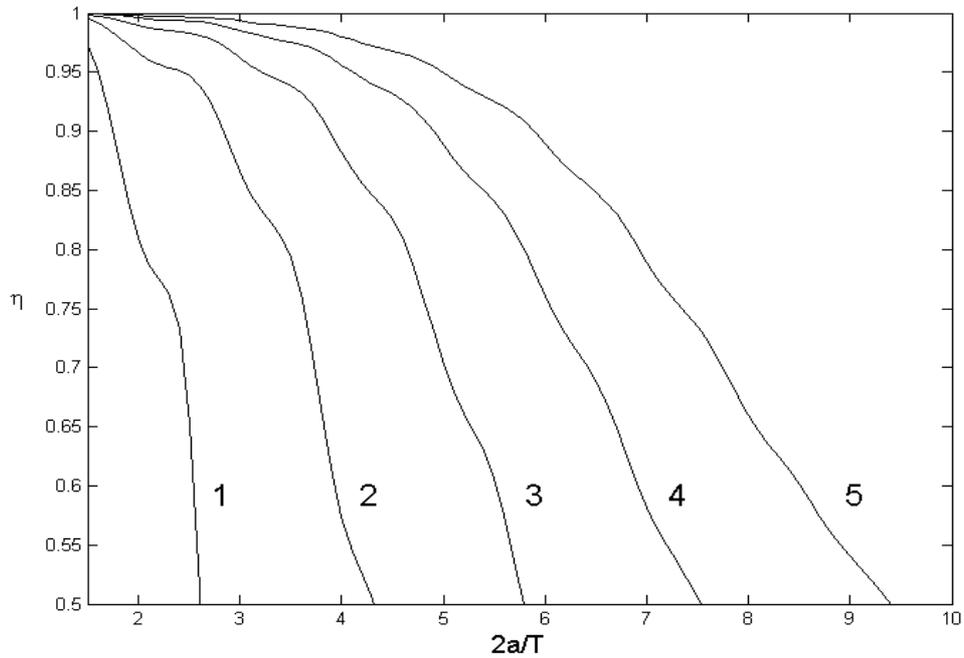

. Fig. 7- The degree of periodicity as a function of the sampling interval in units of the period. The numbers 1-5 correspond to $K-1$ $(N=3K)$.

Interestingly the data presented in the figure above can be approximately represented as a scaling function of $\dfrac{a}{N^{1.2}T}$.

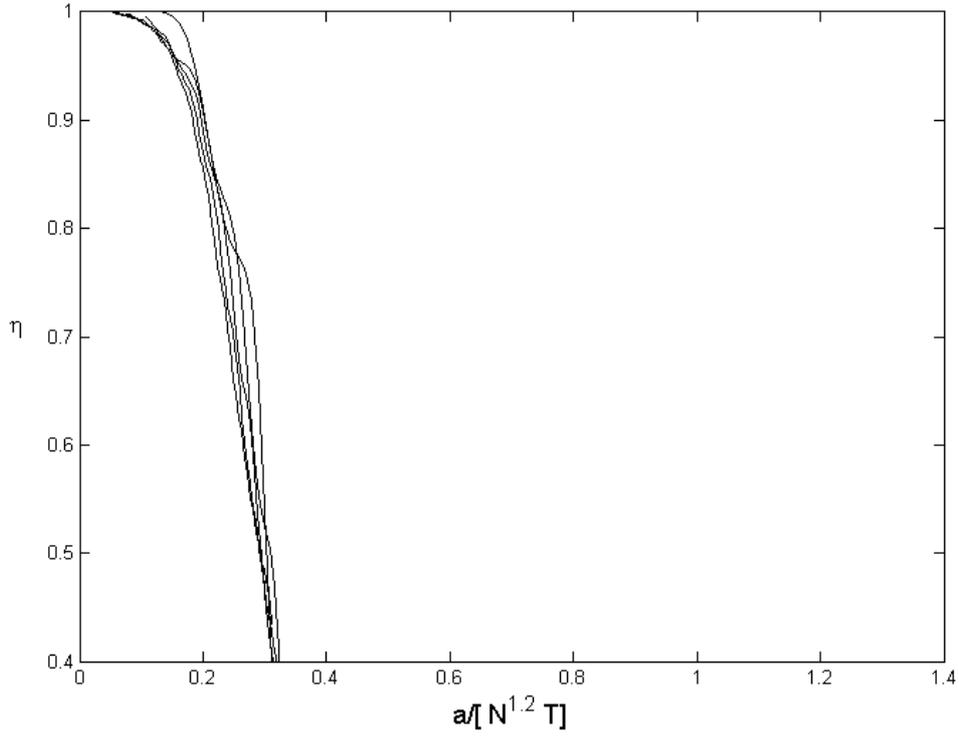

Fig. 8-An approximate scaling representation of the previous graph.

(It remains to be seen, whether this scaling is accidental or generic.)

It is clear, therefore, that to get a larger number of " clean" oscillations we need to increase $M$ while keeping $a$ fixed. Since the ratio between the superoscillation and the band cutoff frequency, $\omega_B$, is given by

$$\frac{\omega}{\omega_B} = \frac{\pi(M-1)}{aN} \qquad (16)$$

and since we use a constant ratio of 2/3 between $M$ and $N$, it means that increasing the superoscillation frequency implies , increasing the number of Fourier components and that means decreasing the superoscillating yield. Thus "clean" oscillations or equivalently a high degree of periodicity come at the price of reduction in the superoscillation yield.

The tradeoff between the two should depend on the actual application at hand. We hope to discuss this point in more detail in future publications.

Bibliography

[1] Y. Aharonov, D. Albert and L. Vaidman, "How the result of a measurement of a component of the spin of a spin-1/2 particle can turn out to be 100", *Phys. Rev. Lett.*, Vol. 60, no. 14, pp. 1351–1354, 1988.

[2] Y. Aharonov, S. Popescu and D. Rohrlich, "How can an infra-red photon behave as a gamma ray?", Tel-Aviv University Preprint TAUP 1847–90, 1990.

[3] M. V. Berry, "Evanescent and real waves in quantum billiards and Gaussian beams", *J. Phys. A: Math. Gen.*, vol. 27, no. 11, pp. L391–L398, 1994.

[4] M. V. Berry, "Faster than Fourier in Quantum Coherence and Reality – Celebration of The 60th Birthday of Yakir Aharonov", ed. J. S. Anandan and J. L. Safko (Singapore: World Scientific) pp. 55–65, 1994.

[5] A. Kempf, "Black holes, bandwidths and Beethoven", *J. Math. Phys.*, vol. 41, no. 4, pp. 2360–2374, Apr. 2000.

[6] A. Kempf and P. J. S. G. Ferreira, "Unusual properties of superoscillating particles", *J. Phys. A: Math. Gen.*, vol. 37, pp. 12067-12076, Dec. 2004.

[7] M. V. Berry and S. Popescu, "Evolution of quantum superoscillations and optical superresolution without evanescent waves", *J. Phys. A: Math. Gen.*, Vol. 39 ,pp. 6965–6977, 2006.

[8] N. I. Zheludov, "What diffraction limit?", *Nature Materials*, vol. 7, pp. 420–422, June 2008.

[9] F. M. Huang and N. I. Zheludov, "Super–Resolution without Evanescent Waves", *Nano Letters*, vol. 9, no 3 , pp. 1249–1254, 2009.

[10] Z. Zalevsky, "Super–Resolved Imaging: Geometrical and Diffraction Approaches", Springer Verlag, Springer Briefs in Physics 2011.

[11] N. I. Zheludov, "The Next Photonic Revolution", *J. Opt. A: Pure Appl"*, *Opt.*,vol. 11, 110202, 2011.

[12] A.M.H. Wong and G.V. Eleftheriades, "Superoscillatory Radar Imaging: Improving Radar Range Resolution Beyond Fundamental Bandwidth Limitations", *IEEE Microwave and Wireless Components Letters*, Vol. 22, no. 3, pp. 147–149, 2012.

[13] A.M.H. Wong and G.V. Eleftheriades, "Adaptation of Schelkunoffs superdirective antenna theory for the realization of superoscillatory antenna

arrays", *IEEE Antennas Wireless Propag. Lett.*, vol. 9, pp. 315318, Apr. 2010
.

[14] D. Slepian and H. O. Pollak, "Prolate spheroidal wave functions, Fourier analysis and uncertainty–I", *Bell Syst. Tech. J.*, vol. 40, no. 1, pp. 43–63, Jan. 1961.

[15] L. Levi, "Fitting a bandlimited signal to given points", *IEEE Trans. Inf. Theory*, vol. IT–11, pp. 372376, Jul. 1965.

[16] D. Slepian, "Prolate spheroidal wave functions, Fourier analysis and uncertainty–V: The discrete case", *Bell Syst. Tech. J.*, vol. 57, no. 5, pp. 1371–1430, May 1978.

[17] P. J. S. G. Ferreira and A. Kempf, "The energy expense of superoscillations",
in Signal Process. XITheories Applicat.: Proc. EUSIPCO-2002
XI Eur. Signal Process. Conf., Toulouse, France, Sep. 2002, vol. II, pp. 347–350.

[18] P. J. S. G. Ferreira and A. Kempf, "Superoscillations: Faster Than the Nyquist Rate", *IEEE Trans. on Signal Processing*, vol. 54, no. 10, pp.

[19] P. J. S. G. Ferreira, "Stable interpolation and error correction: concatenated real-codes with an interleaver" in: Proceedings of the 2001 Workshop on Sampling Theory and Applications, SampTA'2001, Orlando, Florida, pp. 177–182, 2001.

[20] P. J. S. G. Ferreira, "Stability Issues in Error Control Coding in the Complex Field, Interpolation, and Frame Bounds". *IEEE Signal Processing Letters*, Vol. 7, no. 3, pp. 57–59, 2000.

[21] P. J. S. G. Ferreira, "The stability of a procedure for the recovery of lost samples in band-limited signals", *Signal Process.*, Vol. 40, no. 3, pp. 195-205, 1994.

[22] E. Katzav and M. Schwartz, "Yield Optimized Superoscillations",IEEE Trans. on Signal Proc. Vol. 61,no 12, pp 3113-3118, 2013.